\documentclass[journal=jacsat,manuscript=article]{achemso}
\usepackage[T1]{fontenc} 
\usepackage{float}
\usepackage{epstopdf}
\usepackage{indentfirst}

\usepackage{titlesec}
\usepackage{amsmath}
\usepackage{graphicx}
\usepackage{subfigure}
\usepackage[version=3]{mhchem}

 \newcommand{\Rmnum}[1]{\expandafter\@slowromancap\romannumeral #1@} \makeatother

\titleformat{\section}[block]{\large\bfseries}{\Roman{section}.}{1em}{}
\titleformat{\subsection}[block]{\large\bfseries}{\Alph{subsection}.}{1em}{}

\author{Baopi Liu}
\email{bpliu@mail.bnu.edu.cn}
\affiliation{Department of Physics, Beijing Normal University, Beijing, 100875, China}

\title[An \textsf{achemso} demo]
      {Spherical-harmonic Expansion of the Modified Diffusion Equation for Wormlike Chain in Curvilinear Coordinates}

\abbreviations{IR,NMR,UV}

\begin{document}
\begin{abstract}
We investigate the wormlike polymer chains using self-consistent field theory and take into account the Onsager excluded-volume interaction between polymer segments. The propagator of polymer chain is one of the essential physical quantities used to study the conformation of polymers, which satisfies the modified diffusion equation (MDE) for wormlike chain. The propagator of wormlike chain is not only dependent on the spatial variables, but also on the orientation. We separate the variables of propagator by using spherical-harmonic series and then simplify the MDE to a coupled set of equations only depends on spatial variables in this paper. We expand the MDE by spherical-harmonic functions in cylindrical coordinates and spherical coordinates, respectively. We find that there are three ways to set the orientation, no matter in cylindrical coordinates or spherical coordinates. But for the convenience of calculation, we compare these three forms and choose the simplest one to simplify the MDE. And we get a coupled set of equations only depends on spatial variables.
\end{abstract}

\newpage
\section{Introduction}
\begin{figure}[H]
 \centering
     \includegraphics[scale=0.60]{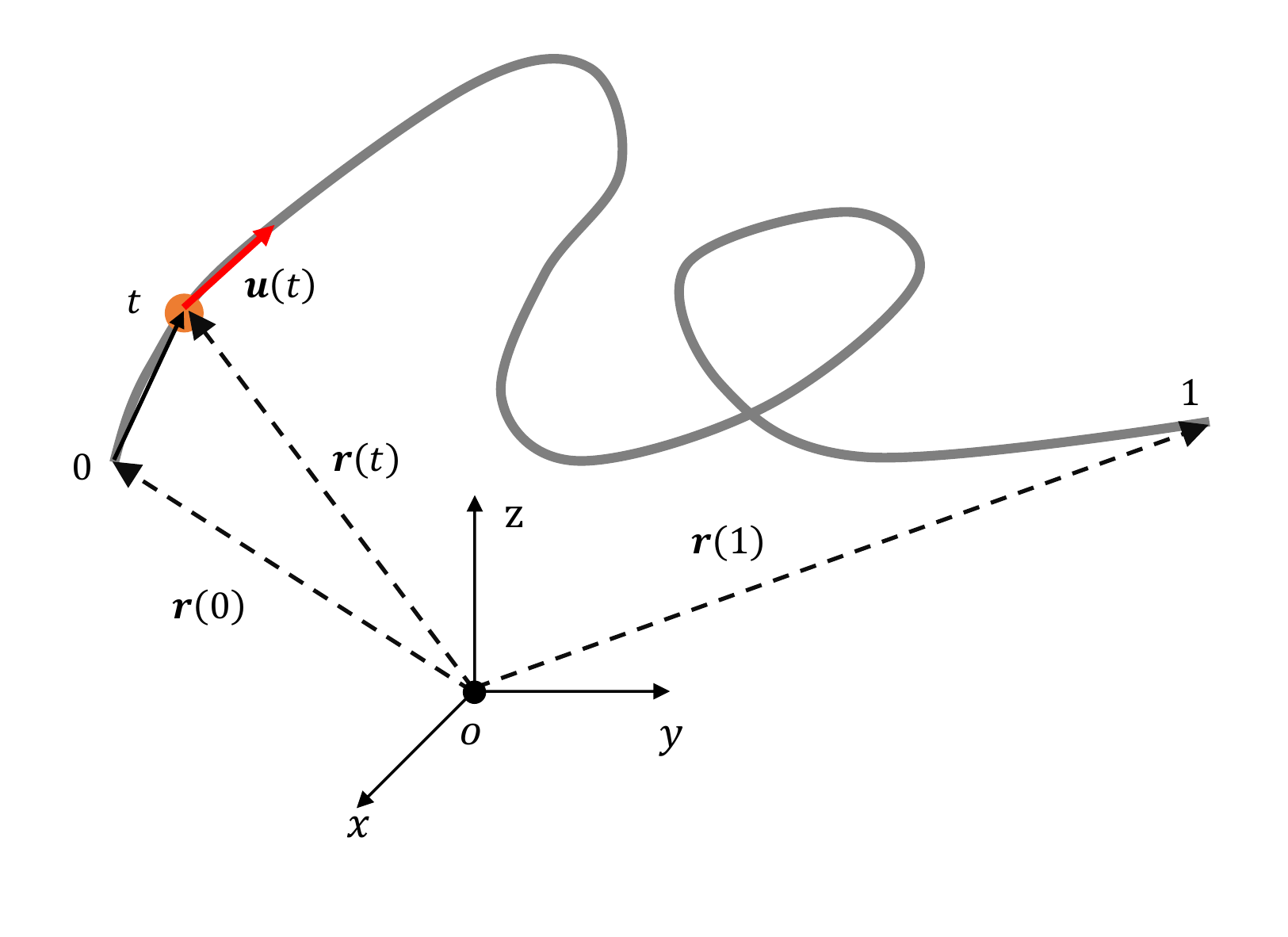}
     \caption{The continuous wormlike chain model describes the configuration of a polymer as a space curve~$\textbf{r}(t)$, where~$t\in [0,1]$~is a contour parameter. The chain end positions correspond to~$\textbf{r}(0)$~and~$\textbf{r}(1)$.}
     \label{fig:1}
\end{figure}

Semiflexible polymer chains have been widely used to describe the structures and dynamics of a large variety of synthetic and biological polymers such as liquid crystalline polymers and DNA molecule. Kratky and Porod~\cite{1} introduced the wormlike chain model to describe semiflexible polymer chains. In Saito-Takahashi-Yunoki (STY)~\cite{2} treatment, the configuration of the continuous wormlike chain is specified by a space curve~$\textbf{r}(t)$ in which $t \in [0,1]$~is a contour variable that describes the location of a segment along the backbone of the chain, as shown in Figure~\ref{fig:1}. The vector~$\textbf{u} (t) \equiv d \textbf{r} (t)/Ldt$~is a tangent vector to the chain at contour location~$t$~and is constrained to be a unit vector~\cite{3,4}, where~$L$~is the contour length.

In the external field~$W(\textbf{r}, \textbf{u})$, the propagator~$q(\textbf{r}, \textbf{u}; t)$~of wormlike chains satisfies the modified diffusion equation (MDE)~\cite{4}. Onsager~\cite{5} developed the polymer chains interact with each other through the excluded-volume interaction, as a function of the density distribution. It is interesting to study the polymers confined in a restricted space, either inside a cylindrical~\cite{9,10,12,13,15} or spherical pore~\cite{16,17}. In these cases the usage of curvilinear coordinates for representing $\textbf{r}(t)$ is more convenient than the Cartesian coordinates. Theoretical understanding of polymer chains in confinement, is a topic that has been widely studied~\cite{6,7,8,11,14,18,19,20,21}. Previous works on wormlike chains have been carried out in Cartesian coordinates~\cite{19,20,21,22,23}. However, in Qin Liang's work~\cite{24}, they demonstrated that the mathematical form of MDE in curvilinear coordinates is quite different from that in Cartesian coordinates, and an additional term in the partial differential equation needs to be accounted in curvilinear coordinates.

In this paper, we expand the MDE using spherical-harmonic series~\cite{23} in cylindrical and spherical coordinates. In this process, the most difficulty is how to match the derivative terms so that they can be expressed by the product of three spherical harmonics, and then expressed as the Clebsch-Gordan coefficients. For convenience to proceed our calculation, it is significant to choose appropriate form of tangent vector~$\textbf{u}$.

\section{Model and Theory}
\hangafter=-1\hangindent=19pt\noindent

We consider a monodisperse solution of~$n$~wormlike chains in a structureless solvent, occupying a total volume~$V$. Each chain is characterized by persistence length~$\lambda$~and its cross-section diameter~$d$ ($L, \lambda \gg d$). The propagator satisfies the MDE~\cite{24}:
\begin{equation}\label{MDE2}
\frac{\partial}{\partial t}q(\textbf{r}, \textbf{u}; t)=\left\{-L\textbf{u}\cdot \nabla_{\textbf{r}}+\frac{L}{2 \lambda} \nabla^2_{\textbf{u}}-W(\textbf{r}, \textbf{u})
+L[(\textbf{u} \cdot \nabla_{\textbf{r}})\textbf{u}] \cdot \nabla_{\textbf{u}}\right\} q(\textbf{r}, \textbf{u}; t)
\end{equation}
\begin{equation}\label{MDE3}
\frac{\partial}{\partial t}q^{\ast}(\textbf{r}, \textbf{u}; t)=\left\{-L\textbf{u}\cdot \nabla_{\textbf{r}}-\frac{L}{2 \lambda} \nabla^2_{\textbf{u}}+W(\textbf{r}, \textbf{u})
-L[(\textbf{u} \cdot \nabla_{\textbf{r}})\textbf{u}] \cdot \nabla_{-\textbf{u}}\right\} q^{\ast}(\textbf{r}, \textbf{u}; t)
\end{equation}
Since the two ends of the polymer chains are different, we introduce the function~$q^{*}(\textbf{r}, \textbf{u}; t)$, which is defined as the propagator starting from the opposite end of the polymer, and~$q^{\ast}(\textbf{r}, \textbf{u}; t)=q(\textbf{r}, -\textbf{u}; 1-t)$~with initial conditions
\begin{equation}
q(\textbf{r}, \textbf{u}; 0)=1
\end{equation}
\begin{equation}
q^{\ast}(\textbf{r}, \textbf{u}; 1)=1
\end{equation}

The free energy of the system is~\cite{18}
\begin{equation}\label{eq:freenergy}
\beta F=da^{2}\int d\textbf{r}d\textbf{u}\int d\textbf{u}'\rho(\textbf{r}, \textbf{u})|\textbf{u}\times \textbf{u}'|\rho(\textbf{r}, \textbf{u}')-\int d\textbf{r}d\textbf{u}W(\textbf{r}, \textbf{u})\rho(\textbf{r}, \textbf{u})-\ln(\frac{Q^{n}}{n!})
\end{equation}
where~$a$~represents the Kuhn length and is related to the persistence length~$\lambda$~by,~$a\equiv 2\lambda$. And~$Q=\int d\textbf{r}d\textbf{u}q(\textbf{r}, \textbf{u}; t)q^{\ast}(\textbf{r}, \textbf{u}; t)$~is the single chain partition function. Taking the saddle point of the free energy given in equation~(\ref{eq:freenergy})~with respect to~$\rho(\textbf{r}, \textbf{u})$~and~$W(\textbf{r}, \textbf{u})$, we obtain the self-consistent mean-field equations:
\begin{equation}\label{eq:SCFT1}
W(\textbf{r}, \textbf{u})=2da^{2}\int d\textbf{u}'|\textbf{u}\times \textbf{u}'|\rho(\textbf{r}, \textbf{u})
\end{equation}
\begin{equation}\label{eq:SCFT2}
\rho(\textbf{r}, \textbf{u})=\frac{n}{Q}\int_{0}^{1}dtq(\textbf{r}, \textbf{u}; t)q^{\ast}(\textbf{r},\textbf{u}; t)
\end{equation}

Expanding~$\rho(\textbf{r}, \textbf{u})$, $W(\textbf{r}, \textbf{u})$, $q(\textbf{r}, \textbf{u}; t)$~and~$q^{\ast}(\textbf{r}, \textbf{u}; t)$~in terms of spherical-harmonic series~\cite{23}:
\begin{eqnarray}
\begin{aligned}
      &\rho(\textbf{r}, \textbf{u})=\sum\limits_{l,m}\rho_{l,m}(\textbf{r})Y_{l,m}(\textbf{u}),\qquad
      W(\textbf{r}, \textbf{u})=\sum\limits_{l,m}W_{l,m}(\textbf{r})Y_{l,m}(\textbf{u})\\
      &q(\textbf{r}, \textbf{u}; t)=\sum\limits_{l,m}q_{l,m}(\textbf{r},t)Y_{l,m}(\textbf{u}),\qquad
      q^{\ast}(\textbf{r}, \textbf{u}; t)=\sum\limits_{l,m}q_{l,m}^{\ast}(\textbf{r},t)Y_{l,m}(\textbf{u})
\end{aligned}
\end{eqnarray}
Since the research objects are all real functions, the expansion coefficients must obey the following conditions~\cite{23}:
\begin{equation}
\rho_{l,m}(\textbf{r})=(-1)^{m}\rho_{l,-m}^{\ast}(\textbf{r})
\end{equation}
Next, the spherical-harmonic expansion for the kernel~$|\textbf{u}\times \textbf{u}{'}|$~by use of the addition theorem~\cite{18}
\begin{equation}
|\textbf{u}\times \textbf{u}{'}|=\sum\limits_{l,m}\frac{4\pi}{2l+1}d_{l}\cdot Y_{l,m}(\textbf{u})Y_{l,m}^{\ast}(\textbf{u}{'})
\end{equation}
with~\cite{18}
\begin{equation}
d_{l}=
\begin{cases}
     \displaystyle \frac{\pi}{4}, & l=0,\\
     \displaystyle 0, & l=1,3,5,\dots,\\
     \displaystyle -\frac{\pi(2l+1)l!(l-2)!}{2^{2l+1}(\frac{l}{2}-1)!(\frac{l}{2})!(\frac{l}{2})!(\frac{l}{2}+1)!}, & l=2,4,6,\dots
\end{cases}
\end{equation}

The mean-field equations~(\ref{eq:SCFT1})~and~(\ref{eq:SCFT2})~can be rewrote as
\begin{equation}
W_{l,m}(\textbf{r})=\frac{8\pi}{2l+1}d_{l}\cdot da^{2}\rho_{l,m}(\textbf{r})
\end{equation}
\begin{equation}\label{eq:trigonometric}
\begin{split}
\rho_{l,m}(\textbf{r})&=\frac{n}{Q}\int_{0}^{1}dt\sum\limits_{l{'},m{'}}\sum\limits_{l{''},m{''}}q_{l{'},m{'}}^{\ast}(\textbf{r},t)q_{l{''},m{''}}(\textbf{r},t)\int
  d\textbf{u}Y_{l{'},m{'}}(\textbf{u})Y_{l{''},m{''}}(\textbf{u})Y_{l,m}^{\ast}(\textbf{u})\\
  &=\frac{n}{Q}\int_{0}^{1}dt\sum\limits_{l{'},m{'}}\sum\limits_{l{''},m{''}}q_{l{'},m{'}}^{\ast}(\textbf{r},t)q_{l{''},m{''}}(\textbf{r},t)
  \sqrt{\frac{(2l{''}+1)(2l{'}+1)}{4\pi(2l+1)}}\times C_{0,0,0}^{l{''},l{'},l}C_{m{''},m{'},m}^{l{''},l{'},l}
\end{split}
\end{equation}
where~$C_{m{''},m{'},m}^{l{''},l{'},l}$~are Clebsch-Gordan coefficients, it is the result of integral for three spherical harmonics.

\begin{equation}
q_{l,m}(\textbf{r},t)=\int d\textbf{u}Y_{l,m}^{\ast}(\textbf{u})q(\textbf{r}, \textbf{u}; t)
\end{equation}
Equation~(\ref{MDE2})~becomes:
\begin{equation}\label{eq:MDE1}
\begin{split}
\frac{\partial}{\partial t}q_{l,m}(\textbf{r},t)&=\int d\textbf{u}Y_{l,m}^{\ast}(\textbf{u})\frac{\partial}{\partial t}q(\textbf{r}, \textbf{u}; t)\\
&=\int d\textbf{u}Y_{l,m}^{\ast}(\textbf{u})\left\{-L\textbf{u}\cdot \nabla_{\textbf{r}}+\frac{L}{2 \lambda} \nabla^2_{\textbf{u}}-W(\textbf{r}, \textbf{u})
+L[(\textbf{u} \cdot \nabla_{\textbf{r}})\textbf{u}] \cdot \nabla_{\textbf{u}}\right\}\sum\limits_{l{'},m{'}}q_{l{'},m{'}}(\textbf{r},t)Y_{l{'},m{'}}(\textbf{u})
\end{split}
\end{equation}

\section{Spherical-harmonic expansion}

\subsection{Cylindrical coordinates}

\begin{figure}[H]
 \centering
     \includegraphics[scale=0.90]{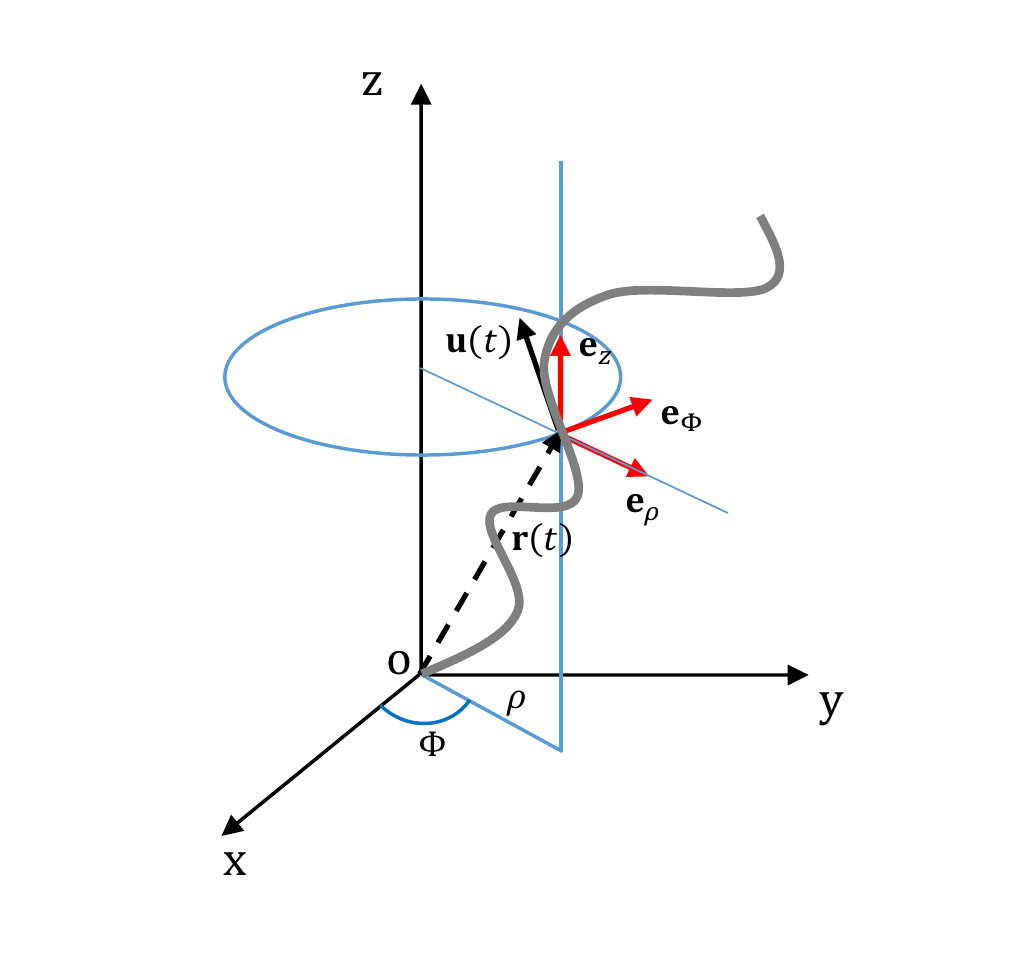}
     \caption{The considered polymer is represented by gray line, the spatial variable~$\textbf{r}$~is represented by variables~$\rho$, $\Phi$, $z$~in cylindrical coordinates. The local coordinate frame is given by orthonormal unit vectors~$\textbf{e}_{\rho}$,~$\textbf{e}_{\Phi}$, and~$\textbf{e}_{z}$.}
     \label{fig:2}
\end{figure}

In the cylindrical coordinate system, the spatial variable~$\textbf{r}$~is represented by variables~$\rho$, $\Phi$, $z$, the unit vector are given by the orthonormal unit vectors~$\textbf{e}_{\rho}$,~$\textbf{e}_{\Phi}$, $\textbf{e}_{z}$, as shown in Figure~\ref{fig:2}.

\begin{figure}[H]
 \centering
     \includegraphics[scale=0.70]{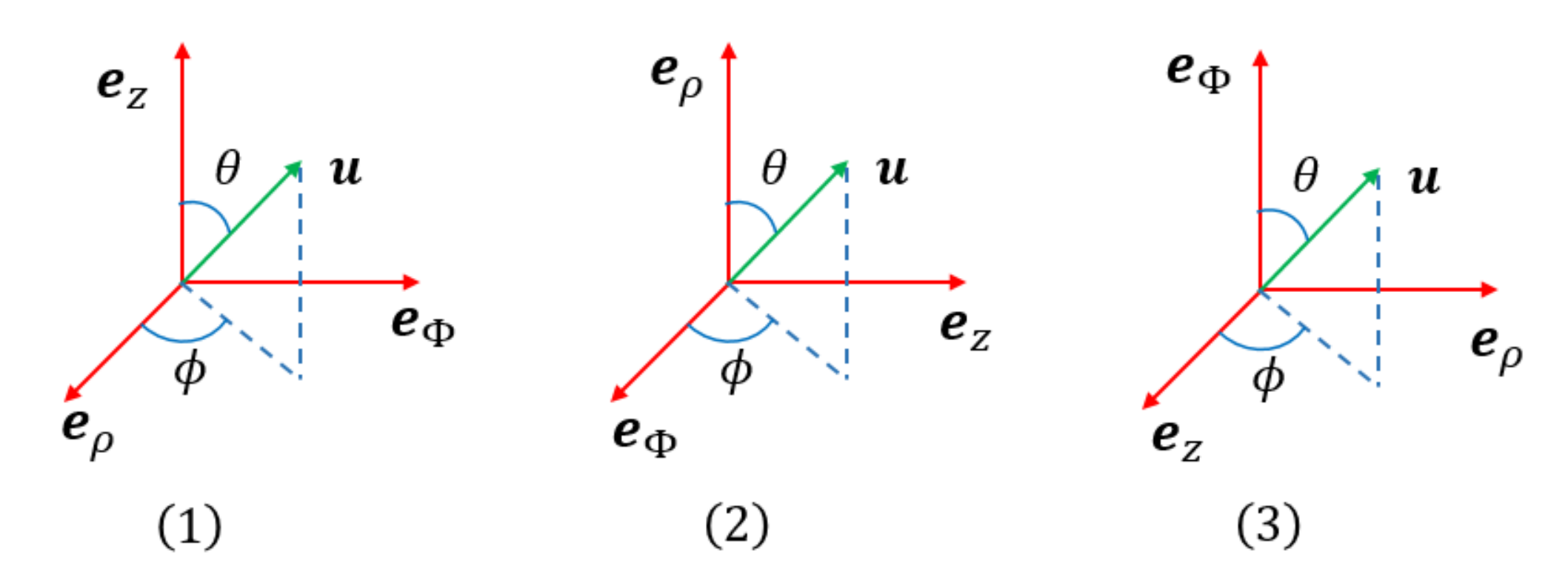}
     \caption{In the cylindrical coordinate system, the three forms of the orientation vector~$\textbf{u}$.}
     \label{fig:3}
\end{figure}

We establish a spherical coordinate system for~$\textbf{u}$~within this local frame~\cite{24}, and there are three forms of~$\textbf{u}$, as shown in Figure~\ref{fig:3}.
For calculation convenience, we set the orientation vector~$\textbf{u}$~in the following form~(Figure~\ref{fig:3}(1))
\begin{equation}
\textbf{u}=\sin\theta \cos\phi\textbf{e}_{\rho}+\sin\theta \sin\phi\textbf{e}_{\Phi}+\cos\theta\textbf{e}_{z}
\end{equation}
The propagator~$q(\textbf{r}, \textbf{u}; t)$~can be represented by~$q(\rho,\Phi,z,\theta,\phi; t)$, where~$\theta$~is the polar angle and~$\phi$~is the azimuthal angle. And the derivative terms in equation~(\ref{eq:MDE1})~become
\begin{equation}
\textbf{u}\cdot \nabla_{\textbf{r}}q(\textbf{r}, \textbf{u}; t)=\Big(\sin\theta \cos\phi \frac{\partial}{\partial \rho}+\frac{\sin\theta \sin\phi}{\rho}\frac{\partial}{\partial \Phi}+\cos\theta\frac{\partial}{\partial z}\Big)q(\textbf{r}, \textbf{u}; t)
\end{equation}
and
\begin{equation}\label{eq:de2}
[(\textbf{u}\cdot \nabla_{\textbf{r}})\textbf{u}]\cdot \nabla_{\textbf{u}} q(\textbf{r}, \textbf{u}; t)=\frac{\sin\theta \sin\phi}{\rho}\frac{\partial}{\partial \phi}q(\textbf{r}, \textbf{u}; t)
\end{equation}
The last term in the right of equation~(\ref{eq:MDE1})~is zero in Cartesian coordinates, but as can be seen from equation~(\ref{eq:de2}), it is non-zero in cylindrical coordinates.

Expanding equation~(\ref{eq:MDE1})~using spherical-harmonic series, we obtain the following coupled set of equations in cylindrical coordinates\cite{25}:
\begin{equation}\label{eq:solution1}
\begin{split}
  \frac{\partial}{\partial t}q_{l,m}(\textbf{r},t)=&-L \sum\limits_{l{'},m{'}}\sqrt{\frac{2l{'}+1}{2l+1}}\left\{C_{0,0,0}^{1,l{'},l} \Big[\frac{1}{\sqrt{2}}(C_{-1,m{'},m}^{1,l{'},l}-C_{1,m{'},m}^{1,l{'},l})\frac{\partial}{\partial \rho}\right. \\
  & \left. +\frac{i}{\sqrt{2}}(C_{-1,m{'},m}^{1,l{'},l}+C_{1,m{'},m}^{1,l{'},l})\frac{1}{\rho}\frac{\partial}{\partial \Phi}+C_{0,m{'},m}^{1,l{'},l}\frac{\partial}{\partial z}\Big] \right\}q_{l{'},m{'}}(\textbf{r},t)\\
  &-\frac{L}{2\lambda}l(l+1)q_{l,m}(\textbf{r},t)\\
  &-\sum\limits_{l{'},m{'}}\sum\limits_{l{''},m{''}}\sqrt{\frac{(2l{''}+1)(2l{'}+1)}{4\pi(2l+1)}}C_{0,0,0}^{l{''},l{'},l} \cdot C_{m{''},m{'},m}^{l{''},l{'},l} \cdot W_{l{'},m{'}}(\textbf{r}) \cdot q_{l{''},m{''}}(\textbf{r},t)\\
  &-\sum\limits_{l{'},m{'}}\frac{L}{\sqrt{2}\rho}m{'}\sqrt{\frac{2l{'}+1}{2l+1}}C_{0,0,0}^{1,l{'},l}(C_{-1,m{'},m}^{1,l{'},l}+C_{1,m{'},m}^{1,l{'},l})q_{l{'},m{'}}(\textbf{r},t)
\end{split}
\end{equation}
with initial conditions
\begin{equation}\label{eq:condition1}
q_{0,0}(\textbf{r},0)=\sqrt{4\pi}, \qquad q_{l,m}(\textbf{r},0)=0, \qquad \rm otherwise.
\end{equation}
Similarly, we obtain
\begin{equation}\label{eq:solution2}
\begin{split}
  \frac{\partial}{\partial t}q_{l,m}^{\ast}(\textbf{r},t)=&-L \sum\limits_{l{'},m{'}}\sqrt{\frac{2l{'}+1}{2l+1}}
  \left\{C_{0,0,0}^{1,l{'},l} \Big[\frac{1}{\sqrt{2}}(C_{-1,m{'},m}^{1,l{'},l}-C_{1,m{'},m}^{1,l{'},l})\frac{\partial}{\partial \rho}\right. \\
  & \left. +\frac{i}{\sqrt{2}}(C_{-1,m{'},m}^{1,l{'},l}+C_{1,m{'},m}^{1,l{'},l})\frac{1}{\rho}\frac{\partial}{\partial \Phi}+C_{0,m{'},m}^{1,l{'},l}\frac{\partial}{\partial z}\Big] \right\}q_{l{'},m{'}}^{\ast}(\textbf{r},t)\\
  &+\frac{L}{2\lambda}l(l+1)q_{l,m}^{\ast}(\textbf{r},t)\\
  &+\sum\limits_{l{'},m{'}}\sum\limits_{l{''},m{''}}\sqrt{\frac{(2l{''}+1)(2l{'}+1)}{4\pi(2l+1)}}C_{0,0,0}^{l{''},l{'},l} \cdot C_{m{''},m{'},m}^{l{''},l{'},l} \cdot W_{l{'},m{'}}(\textbf{r}) \cdot q_{l{''},m{''}}^{\ast}(\textbf{r},t)\\
  &-\sum\limits_{l{'},m{'}}\frac{L}{\sqrt{2}\rho}m{'}\sqrt{\frac{2l{'}+1}{2l+1}}C_{0,0,0}^{1,l{'},l}(C_{-1,m{'},m}^{1,l{'},l}+C_{1,m{'},m}^{1,l{'},l})q_{l{'},m{'}}^{\ast}(\textbf{r},t)
\end{split}
\end{equation}
with initial conditions
\begin{equation}\label{eq:condition2}
q_{0,0}^{\ast}(\textbf{r},1)=\sqrt{4\pi}, \qquad q_{l,m}^{\ast}(\textbf{r},1)=0, \qquad \rm otherwise
\end{equation}

According to the above expressions, the single chain partition function and the free energy~(\ref{eq:freenergy})~can be represented by
\begin{equation}
Q=\sum\limits_{l,m}(-1)^{m}\int d\textbf{r}q_{l,m}(\textbf{r},t)q_{l,-m}^{\ast}(\textbf{r},t)
\end{equation}
\begin{equation}
\beta F=-\sum\limits_{l,m}\frac{4\pi}{2l+1}d_{l}\cdot da^{2}\int d\textbf{r}|\rho_{l,m}(\textbf{r})|^{2}-\ln(\frac{Q^{n}}{n!})
\end{equation}

We now analyze the the simplest case where the densities vary in only one spatial dimension, here is chosen to be the~$z$~direction. Then spatial variable of the propagator independent of~$\rho$ and $\Phi$, and the unit vector~$\textbf{u}(t)$~only depends on the polar angle~$\theta$, not on the azimuth angle~$\phi$. Therefor, in equations~(\ref{eq:solution1})~and~(\ref{eq:solution2}), $m=m{'}=m{''}=0$, so that we subsequently drop the subscript which makes the Clebsch-Gordan coefficients $C_{-1,0,0}^{1,l{'},l}=C_{1,0,0}^{1,l{'},l}=0$ and equations~(\ref{eq:solution1})~and~(\ref{eq:solution2})~can be reduced to:
\begin{equation}
\begin{split}
  \frac{\partial}{\partial t}q_{l}(z,t)=
  &-L\sum\limits_{l{'}}\sqrt{\frac{2l{'}+1}{2l+1}}(C_{0,0,0}^{1,l{'},l})^{2}\frac{\partial}{\partial z}q_{l{'}}(z,t)-\frac{L}{2\lambda}l(l+1)q_{l}(z,t)\\
  &-\sum\limits_{l{'},l{''}}\sqrt{\frac{(2l{''}+1)(2l{'}+1)}{4\pi(2l+1)}}(C_{0,0,0}^{1,l{'},l})^{2}\cdot W_{l{'}}(z)\cdot q_{l{''}}(z,t)
\end{split}
\end{equation}
and
\begin{equation}
\begin{split}
  \frac{\partial}{\partial t}q_{l}^{\ast}(z,t)=
  &-L\sum\limits_{l{'}}\sqrt{\frac{2l{'}+1}{2l+1}}(C_{0,0,0}^{1,l{'},l})^{2}\frac{\partial}{\partial z}q_{l{'}}^{\ast}(z,t)-\frac{L}{2\lambda}l(l+1)q_{l}^{\ast}(z,t)\\
  &+\sum\limits_{l{'},l{''}}\sqrt{\frac{(2l{''}+1)(2l{'}+1)}{4\pi(2l+1)}}(C_{0,0,0}^{1,l{'},l})^{2}\cdot W_{l{'}}(z)\cdot q_{l{''}}^{\ast}(z,t)
\end{split}
\end{equation}
with initial conditions
\begin{equation}
q_{0}(z,0)=\sqrt{4\pi}, \qquad q_{l}(z,0)=0, \qquad l>0
\end{equation}
and
\begin{equation}
q_{0}^{\ast}(z,1)=\sqrt{4\pi}, \qquad q_{l}^{\ast}(z,1)=0, \qquad l>0
\end{equation}

Similarly, according to the above expressions, the one-dimensional single-chain partition function and the free energy in equation~(\ref{eq:freenergy})~can be rewritten as
\begin{equation}
Q=\sum\limits_{l}\int dz q_{l}(z,t)q_{l}^{\ast}(z,t)
\end{equation}
\begin{equation}
\beta F=-\sum\limits_{l}\frac{4\pi}{2l+1}d_{l}\cdot da^{2}\int dz \rho_{l}^{2}(z)-\ln(\frac{Q^{n}}{n!})
\end{equation}
This result is consistent with that obtained by Dominik Duchs et al.~\cite{23}.

\subsection{Spherical coordinates}
\begin{figure}[H]
 \centering
     \includegraphics[scale=0.90]{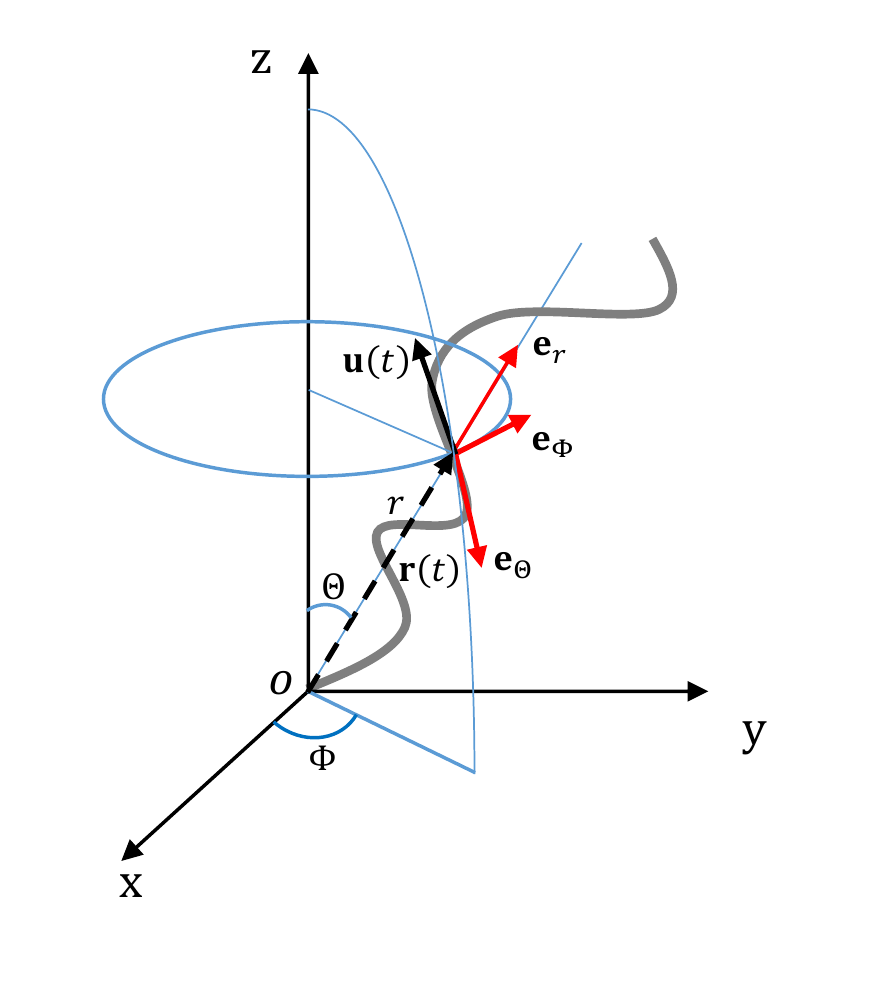}
     \caption{The considered polymer is represented by gray line, the spatial variable~$\textbf{r}$~is represented by variables~$\Theta$, $\Phi$, $r$~in spherical coordinates. The local coordinate frame is given by orthonormal unit vectors~$\textbf{e}_{\Theta}$, $\textbf{e}_{\Phi}$, and~$\textbf{e}_{r}$.}
     \label{fig:4}
\end{figure}

\begin{figure}[H]
 \centering
     \includegraphics[scale=0.70]{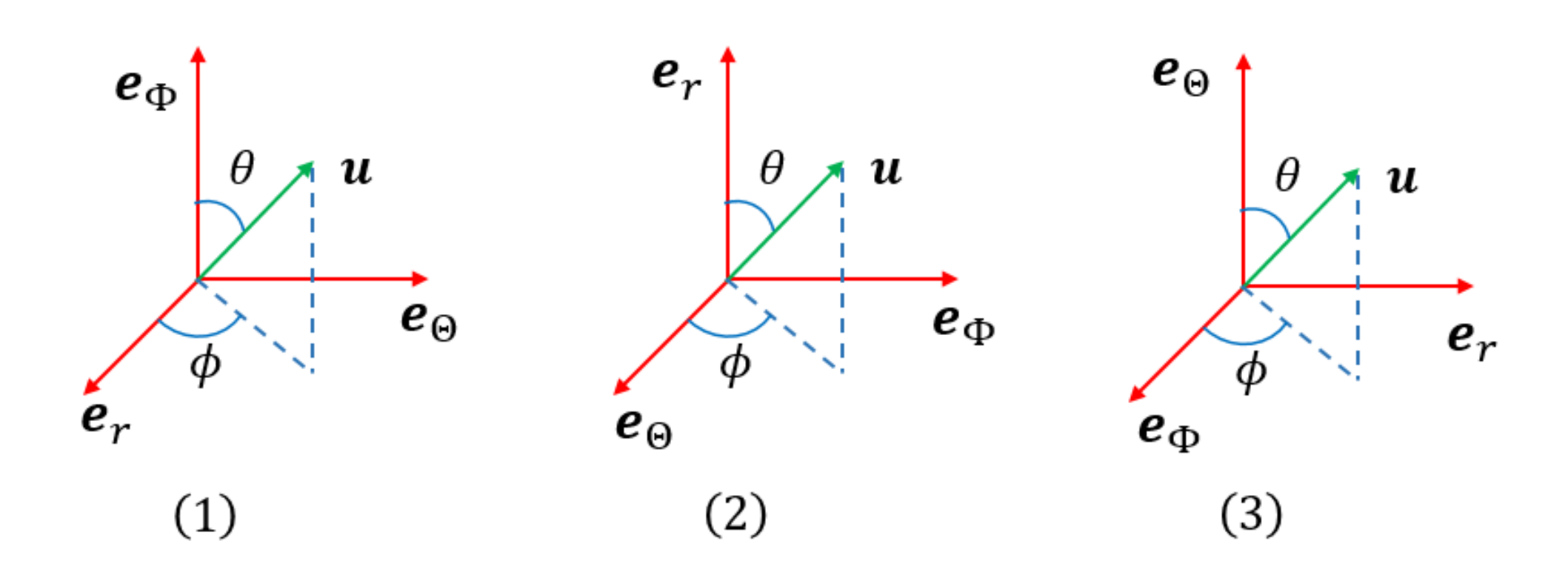}
     \caption{In the spherical coordinate system, the three forms of the orientation vector~$\textbf{u}$.}
     \label{fig:5}
\end{figure}

In the spherical coordinate system, the spatial variable~$\textbf{r}$~is represented by variables~$\Theta$, $\Phi$, $r$, the unit vector are given by the orthonormal unit vectors~$\textbf{e}_{\Theta}$,~$\textbf{e}_{\Phi}$,~$\textbf{e}_{r}$, as shown in Figure~\ref{fig:4}.
We establish a spherical coordinate system for $\textbf{u}$ within this local frame~\cite{24}, and there are three forms of $\textbf{u}$, as shown in Figure~\ref{fig:5}~.
For calculation convenience, we set the orientation vector~$\textbf{u}$ in the following form~(Figure~\ref{fig:5}(2))
\begin{equation}
\textbf{u}=\sin\theta \cos\phi\textbf{e}_{\Theta}+\sin\theta \sin\phi\textbf{e}_{\Phi}+\cos\theta\textbf{e}_{r}
\end{equation}
The propagator~$q(\textbf{r}, \textbf{u}; t)$~is represented by~$q(\Theta, \Phi, r, \theta, \phi; t)$, where~$\theta$~is the polar angle and~$\phi$~is the azimuthal angle. And the derivative terms in equation $(15)$ can be rewritten as
\begin{equation}
\textbf{u}\cdot \nabla_{\textbf{r}}q(\textbf{r}, \textbf{u}; t)=\Big(\frac {\sin\theta \cos\phi}{r} \frac{\partial}{\partial \Theta}+\frac{\sin\theta \sin\phi}{r \sin\Theta}\frac{\partial}{\partial \Phi}+\cos\theta\frac{\partial}{\partial r}\Big)q(\textbf{r}, \textbf{u}; t)
\end{equation}
and
\begin{equation}
[(\textbf{u}\cdot \nabla_{\textbf{r}})\textbf{u}]\cdot \nabla_{\textbf{u}} q(\textbf{r}, \textbf{u}; t)=\Big(\frac{\sin\theta}{r}\frac{\partial}{\partial \theta}+\frac{\sin \theta \sin\phi \cos\Theta}{r \sin\Theta}\frac{\partial}{\partial \phi}\Big)q(\textbf{r}, \textbf{u}; t)
\end{equation}

Expanding equation~(\ref{eq:MDE1})~using spherical-harmonic series, yields the following coupled set of equations in spherical coordinates~\cite{25}:
\begin{equation}
\begin{split}
  \frac{\partial}{\partial t}q_{l,m}(\textbf{r},t)=&-L \sum\limits_{l{'},m{'}}\sqrt{\frac{2l{'}+1}{2l+1}}\left\{C_{0,0,0}^{1,l{'},l} \Big[\frac{1}{\sqrt{2}}(C_{-1,m{'},m}^{1,l{'},l}-C_{1,m{'},m}^{1,l{'},l})\frac{1}{r}\frac{\partial}{\partial \Theta}\right. \\
  & \left. +\frac{i}{\sqrt{2}}(C_{-1,m{'},m}^{1,l{'},l}+C_{1,m{'},m}^{1,l{'},l})\frac{1}{r \sin\Theta}\frac{\partial}{\partial \Phi}+C_{0,m{'},m}^{1,l{'},l}\frac{\partial}{\partial r}\Big] \right\}q_{l{'},m{'}}(\textbf{r},t)\\
  &-\frac{L}{2\lambda}l(l+1)q_{l,m}(\textbf{r},t)\\
  &-\sum\limits_{l{'},m{'}}\sum\limits_{l{''},m{''}}\sqrt{\frac{(2l{''}+1)(2l{'}+1)}{4\pi(2l+1)}}C_{0,0,0}^{l{''},l{'},l} \cdot C_{m{''},m{'},m}^{l{''},l{'},l} \cdot W_{l{'},m{'}}(\textbf{r}) \cdot q_{l{''},m{''}}(\textbf{r},t)\\
  &+\sum\limits_{l{'},m{'}}\frac{L}{\sqrt{2} r}\sqrt{\frac{2l{'}+1}{2l+1}}C_{0,0,0}^{1,l{'},l}\Big[\sqrt{(l{'}-m{'})(l{'}+m{'}+1)}C_{-1,m{'}+1,m}^{1,l{'},l}\\
  &-\sqrt{(l{'}+m{'})(l{'}-m{'}+1)}C_{1,m{'}-1,m}^{1,l{'},l}-\frac{\cos \Theta}{\sin \Theta}m{'}(C_{-1,m{'},m}^{1,l{'},l}+C_{1,m{'},m}^{1,l{'},l})\Big]q_{l{'},m{'}}(\textbf{r},t)
\end{split}
\end{equation}
The initial conditions are
\begin{equation}
q_{0,0}(\textbf{r},0)=\sqrt{4\pi}, \qquad q_{l,m}(\textbf{r},0)=0,  \qquad \rm otherwise
\end{equation}
Similarly, we obtain
\begin{equation}
\begin{split}
  \frac{\partial}{\partial t}q_{l,m}^{\ast}(\textbf{r},t)=&L \sum\limits_{l{'},m{'}}\sqrt{\frac{2l{'}+1}{2l+1}}\left\{C_{0,0,0}^{1,l{'},l} \Big[\frac{1}{\sqrt{2}}(C_{-1,m{'},m}^{1,l{'},l}-C_{1,m{'},m}^{1,l{'},l})\frac{1}{r}\frac{\partial}{\partial \Theta}\right. \\
  & \left. +\frac{i}{\sqrt{2}}(C_{-1,m{'},m}^{1,l{'},l}+C_{1,m{'},m}^{1,l{'},l})\frac{1}{r \sin\Theta}\frac{\partial}{\partial \Phi}+C_{0,m{'},m}^{1,l{'},l}\frac{\partial}{\partial r}\Big] \right\}q_{l{'},m{'}}^{\ast}(\textbf{r},t)\\
  &-\frac{L}{2\lambda}l(l+1)q_{l,m}^{\ast}(\textbf{r},t)\\
  &-\sum\limits_{l{'},m{'}}\sum\limits_{l{''},m{''}}\sqrt{\frac{(2l{''}+1)(2l{'}+1)}{4\pi(2l+1)}}C_{0,0,0}^{l{''},l{'},l} \cdot C_{m{''},m{'},m}^{l{''},l{'},l} \cdot W_{l{'},m{'}}(\textbf{r}) \cdot q_{l{''},m{''}}^{\ast}(\textbf{r},t)\\
  &+\sum\limits_{l{'},m{'}}\frac{L}{\sqrt{2} r}\sqrt{\frac{2l{'}+1}{2l+1}}C_{0,0,0}^{1,l{'},l}\Big[\sqrt{(l{'}-m{'})(l{'}+m{'}+1)}C_{-1,m{'}+1,m}^{1,l{'},l}\\
  &-\sqrt{(l{'}+m{'})(l{'}-m{'}+1)}C_{1,m{'}-1,m}^{1,l{'},l}+\frac{\cos \Theta}{\sin \Theta}m{'}(C_{-1,m{'},m}^{1,l{'},l}+C_{1,m{'},m}^{1,l{'},l})\Big]q_{l{'},m{'}}^{\ast}(\textbf{r},t)
\end{split}
\end{equation}
with initial conditions
\begin{equation}
q_{0,0}^{\ast}(\textbf{r},1)=\sqrt{4\pi}, \qquad q_{l,m}^{\ast}(\textbf{r},1)=0, \qquad \rm otherwise
\end{equation}

According to the above expressions, the single chain partition function and the free energy~(\ref{eq:freenergy})~can be rewritten as
\begin{equation}
Q=\sum\limits_{l,m}(-1)^{m}\int d\textbf{r}q_{l,m}(\textbf{r},t)q_{l,-m}^{\ast}(\textbf{r},t)
\end{equation}
\begin{equation}
\beta F=-\sum\limits_{l,m}\frac{4\pi}{2l+1}d_{l}\cdot da^{2}\int d\textbf{r}|\rho_{l,m}(\textbf{r})|^{2}-\ln(\frac{Q^{n}}{n!})
\end{equation}

\section{Results and Discussion}
In this paper, using wormlike chain model to describe the semiflexible polymers with the Onsager excluded-volume interaction, we are able to obtain the MDE in an external field. Taking the saddle point approximation of the free energy function with $\rho(\textbf{r}, \textbf{u})$ and $W(\textbf{r}, \textbf{u})$, we obtained the mean-field equations. It is more convenient to represent~$\textbf{r}$~in curvilinear coordinates when wormlike chains confined in a cylindrical or spherical pore. The solution of the propagator in an external field is an essential tool to calculate other properties such as the average polymer conformation, the density profile and the free energy of a spatially and orientational inhomogeneous system. In order to obtain the solution of the propagator, we expanded the MDE using spherical-harmonic series. And we obtained the expansion of spherical-harmonic series of MDE in cylindrical and spherical coordinates.

We find that there are three ways to set the orientation vector~$\textbf{u}$~in cylindrical coordinates and spherical coordinates, respectively. Compare the three forms of orientation vector~$\textbf{u}$~, we choose the appropriate one to make the expansion form most simplest. The most difficulty of this investigation is that the gradient which depend on the spatial variable~$\textbf{r}$~and the gradient, divergence and Laplace operate which depend on the orientation~$\textbf{u}$~are mixed together. We expressed the terms which depend on the orientational variable~$\textbf{u}$~as the product of three spherical harmonics. Then we simplify the MDE to a coupled set of equations only depend on the spatial variable. In this paper, we only derive the coupled set of equations but do not do numerical calculation for specific examples. In the following study, we will investigate specific objects using this coupled set of equations.

\paragraph{Acknowledgments:}
This work was supported by the National Natural Science Foundation of China (NSFC) Nos. 21873015, 21434001.

\end{document}